\documentclass[conference]{IEEEtran}
\IEEEoverridecommandlockouts
\usepackage{amsmath,amssymb,amsfonts}
\usepackage{algorithmic}
\usepackage{graphicx}
\usepackage{textcomp}
\usepackage{xcolor}
\usepackage{quantikz}
\usepackage{enumitem}
\usepackage{url}

\def\BibTeX{{\rm B\kern-.05em{\sc i\kern-.025em b}\kern-.08em
    T\kern-.1667em\lower.7ex\hbox{E}\kern-.125emX}}

\usepackage{listings}
\usepackage{xcolor}

\newtheorem{definition}{Definition}
\definecolor{eclipseStrings}{RGB}{42,0.0,255}
\definecolor{eclipseKeywords}{RGB}{127,0,85}
\colorlet{numb}{magenta!60!black}

\let\oldtextbf\textbf
\renewcommand{\textbf}[1]{\oldtextbf{\boldmath #1}}

\lstdefinelanguage{json}{
    basicstyle=\normalfont\ttfamily,
    commentstyle=\color{eclipseStrings}, 
    stringstyle=\color{eclipseKeywords}, 
    numbers=left,
    numberstyle=\scriptsize,
    stepnumber=1,
    numbersep=8pt,
    showstringspaces=false,
    breaklines=true,
    frame=lines,
    string=[s]{"}{"},
    comment=[l]{:\ "},
    morecomment=[l]{:"},
    literate=
        *{0}{{{\color{numb}0}}}{1}
         {1}{{{\color{numb}1}}}{1}
         {2}{{{\color{numb}2}}}{1}
         {3}{{{\color{numb}3}}}{1}
         {4}{{{\color{numb}4}}}{1}
         {5}{{{\color{numb}5}}}{1}
         {6}{{{\color{numb}6}}}{1}
         {7}{{{\color{numb}7}}}{1}
         {8}{{{\color{numb}8}}}{1}
         {9}{{{\color{numb}9}}}{1}
}

\makeatletter
\newcommand{\linebreakand}{%
  \end{@IEEEauthorhalign}
  \hfill\mbox{}\par
  \mbox{}\hfill\begin{@IEEEauthorhalign}
}
\makeatother

\begin{document}

\title{Towards AutoQML: A Cloud-Based Automated Circuit Architecture Search Framework}

\author{\IEEEauthorblockN{Raúl Berganza Gómez}
\IEEEauthorblockA{\textit{Technical University of Munich and} \\ \textit{Data \& Analytics} \\ \textit{E.ON Digital Technology GmbH} \\ Hannover, Germany \\ raul.berganza@tum.de}
\and
\IEEEauthorblockN{Corey O'Meara}
\IEEEauthorblockA{\textit{Data \& Analytics} \\ \textit{E.ON Digital Technology GmbH} \\ Hannover, Germany \\ corey.o'meara@eon.com}
\and
\IEEEauthorblockN{Giorgio Cortiana}
\IEEEauthorblockA{\textit{Data \& Analytics} \\ \textit{E.ON Digital Technology GmbH} \\ Hannover, Germany}
\linebreakand
\IEEEauthorblockN{Christian B. Mendl}
\IEEEauthorblockA{\textit{Department of Informatics} \\ \textit{Technical University of Munich and} \\ \textit{TUM Institute of Advanced Study}\\ Garching, Germany} 
\and
\IEEEauthorblockN{Juan Bernabé-Moreno}
\IEEEauthorblockA{\textit{Data \& Analytics} \\ \textit{E.ON Digital Technology GmbH} \\ Hannover, Germany}
}


\maketitle

\begin{abstract}
The learning process of classical machine learning algorithms is tuned by hyperparameters that need to be customized to best learn and generalize from an input dataset. In recent years, Quantum Machine Learning (QML) has been gaining traction as a possible application of quantum computing which may provide quantum advantage in the future. However, quantum versions of classical machine learning algorithms introduce a plethora of additional parameters and circuit variations that have their own intricacies in being tuned.

In this work, we take the first steps towards Automated Quantum Machine Learning (AutoQML). We propose a concrete description of the problem, and then develop a classical-quantum hybrid cloud architecture that allows for parallelized hyperparameter exploration and model training. 

As an application use-case, we train a quantum Generative Adversarial neural Network (qGAN) to generate energy prices that follow a known historic data distribution. Such a QML model can be used for various applications in the energy economics sector.

\end{abstract}

\begin{IEEEkeywords}
quantum machine learning, parametrized quantum circuit, quantum neural network, software architecture, cloud computing
\end{IEEEkeywords}

\section{Introduction}

Quantum computing entered a new stage of development in recent years, as fundamental breakthroughs and public interest fuel the availability of NISQ devices. Both researchers and industry are working to develop quantum algorithms that demonstrate quantum advantage compared to classical computers while accepting current architectures' constraints. The first step for solving real-world problems with quantum computers is loading data into them. The challenges behind this task are manyfold: the embedding choice affects the possible data transformations, quantum resources are generally scarce and should be accounted for in the embedding choice, and the runtime of existing algorithms for larger datasets due to shared cloud access needs to be considered \cite{sanders_black-box_2019, aaronson_read_2015, schuld_quantum_2021}. Once data has been embedded into qubits, the actual algorithms run on the quantum data. In recent years, many novel quantum algorithms have been proposed that fit into the category of quantum machine learning. These are quantum versions of classical machine learning algorithms such as $k$-NN classification, $k$-means clustering, support vector machine classification, and quantum neural networks \cite{lloyd2013quantum,rebentrost2014quantum,biamonte2017quantum,schuld_supervised_2018,schuld_quantum_2021}.

Unfortunately, even in classical computing, tuning adjustable parameters for each machine learning algorithm can be quite tedious and there exist entire software libraries and frameworks to simplify the process. Some basic examples are parallel multi-core processing for brute-force grid search or parameter optimization techniques \cite{he2018amc,he2021automl,feurer2019auto,feurer2021auto}. These automated tools help machine learning practitioners train and test algorithms. However, they are usually not a 'silver bullet' and the art/skill of the trade is in understanding many aspects of data science such as input feature engineering, proper experimental setup, cross-validation and other techniques. 
In the quantum era of machine learning, quantum data scientists have additional parameters to tune and consider: data embedding type, number of qubits, general Ansatz type for Parametrized Quantum Circuits (PQCs), number of Ansatz layers for PQCs, measurement type, and post-processing techniques. 

In order to overcome the difficulties in long-running quantum machine learning training tasks and determining the best QML model for a given data set, we introduce the first steps towards AutoQML, an automated remote deployment system that runs multiple combinations of hyperparameters to build quantum machine learning models and runs them in parallel in the quantum cloud. The system returns statistics, relevant quantum and machine learning metrics, and the best model with its associated hyperparameter values.

The paper is structured as follows. First, we define the concept of AutoQML and define a hybrid classical-quantum cloud architecture developed to implement it in Section \ref{sec:autoqml}. Following the generic description of the framework, the remaining sections of the paper dive into a specific example of using it for quantum neural network training. Specifically, in Section \ref{sec:example_qnn} we introduce the fundamentals of quantum neural networks and a special subtype called quantum Generative Adversarial Neural Networks (qGANs). We cover some fundamental properties of these machine learning models and discuss certain tunable parameters and the loss function we wish to optimize. We describe an example use-case for training qGANs to learn the best fitting model for an energy economics use-case in Section
\ref{sec:application}. 

The main contributions of the paper are:  we provide the concept and concrete mathematical formulation of the AutoQML problem, and, we define a template for a hybrid cloud architecture which is easily extendable for those wishing to build upon the AutoQML concept. Using the approach, we apply the framework to a sample quantum machine learning problem where we find the best trained qGAN for highly discretized (up to 32 bins) energy price distribution generation. 

\section{AutoQML}\label{sec:autoqml}

In order to discuss the proposed framework, we first provide a precise problem definition that mirrors its classical AutoML counterpart as defined in \cite{feurer2019auto}.

\begin{definition}[AutoQML Problem]
For $i = 1, . . . , n  + m$, let $x_i$ denote a classical data feature
vector and $y_i$ the corresponding classical data target value. Given a training dataset $D_{\text{train}} = \{(x_1, y_1), . . . , ( x_n, y_n)\}$ and the feature vectors $x_{n+1}, . . . , x_{n+m}$ of a test dataset $D_{\text{test}} = \{(x_{n+1}, y_{n+1}), . . . , ( x_{n+m}, y_{n+m})\}$ drawn from the same underlying data
distribution, as well as a classical resource budget $b_{\text{C}}$, a quantum resource budget $b_{\text{Q}}$, and a loss metric $\mathcal{L}(\cdot, \cdot)$, the AutoQML problem is to automatically produce accurate test set predictions $\hat{y}_{n+1}, ... ,  \hat{y}_{n+m}$ using a quantum machine learning model.
The loss of a solution $\hat{y}_{n+1}, . . . ,  \hat{y}_{n+m}$ to the AutoQML problem is given by $\tfrac{1}{m}\sum m_{j=1}^m \mathcal{L}(\hat{y}_{n+j}, y_{n+j})$.
\end{definition}

As in the original definition, the classical resource budget consists of classical memory, CPU clock-time, memory read/write speeds and other constraints. The AutoQML problem encompasses the same classical runtime constraints but also introduces the quantum specific resource budget, which may include factors such as quantum gate operations per second, gate coherence times, qubit counts, and general QPU availability if considering a cloud-based queuing system. 

Given this problem definition, we take the first steps towards an automated quantum machine learning framework. In the approach described in the following subsection, we outline a classical-quantum hybrid cloud architecture, which, with a basic deployment configuration, allows for parallelization of a brute-force approach to hyperparameter search.

\subsection{Distributed Classical-Quantum Architecture}

We propose a cloud-based solution that enables the distributed execution of experiments, results post-processing, parameter benchmarking and plot generation. We illustrate the feasibility of our system with an implementation using Microsoft Azure services. All services used have equivalent routines in other cloud platforms, and porting our solution requires minimal effort due to the few services used. 

Azure Data Factory is an ETL service that allows to integrate and transform data at scale \cite{jonburchel_azure_data_factory}. We use this service due to its orchestration capabilities, its enablement of serverless parallel processing, and its ability to perform synchronization of tasks based on trigger events.

\begin{figure}[t!]
    \centering
    \vspace{-7mm}
    \includegraphics[width=0.98\linewidth]{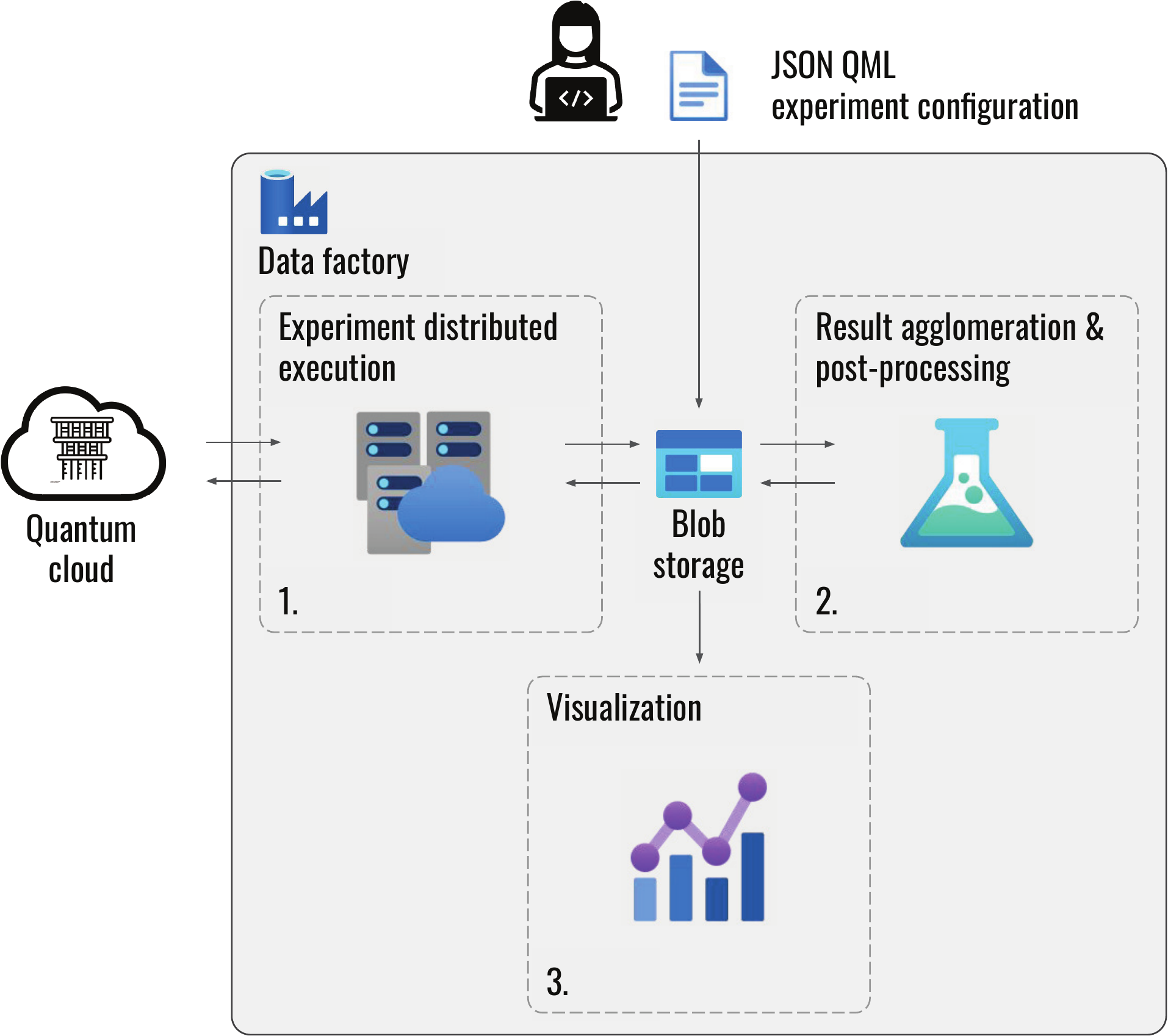}
    \caption{\textbf{Deployment and system architecture of our proposed framework}. The user writes a parameter search configuration file and uploads it to the blob storage assigned to Data Factory, triggering pipeline 1. The pipeline trains QML models asynchronously in a cluster of containers, with each container being able to communicate with quantum backends. Pipeline 2 agglomerates and calculates statistics on the outputs of pipeline 1. Pipeline 3 uses the agglomerated results to generate and persists plots. The three pipelines run sequentially, and the containers communicate through atomic reads and writes to an object store.}
        \vspace{-2mm}
    \label{fig:overall_system_arch}
\end{figure}

Our system, depicted in Fig. \ref{fig:overall_system_arch}, consists of three different pipelines, executed sequentially after the user submits the experiment configuration file: 

\begin{enumerate}[label=\arabic*.]
    \item \textbf{Distributed execution of experiments}: This pipeline creates a pool of worker nodes, distributes the experiments amongst them, and then each node executes and persists its assigned experiments asynchronously. 
    \item \textbf{Agglomeration and post-processing of results}: This pipeline uses a single container to read the files produced by each node, calculate agglomerate statistics, and persist the processed results. This stage also selects the best performing architectural parameters and stores the corresponding model as a pickle file.
    \item \textbf{Plot generation and visualization}: This pipeline uses a single container to read the processed experiment results and generate and persist the plots specified in the configuration file.
\end{enumerate}

\begin{figure}[t]
    \centering
    \includegraphics[width=0.98\linewidth]{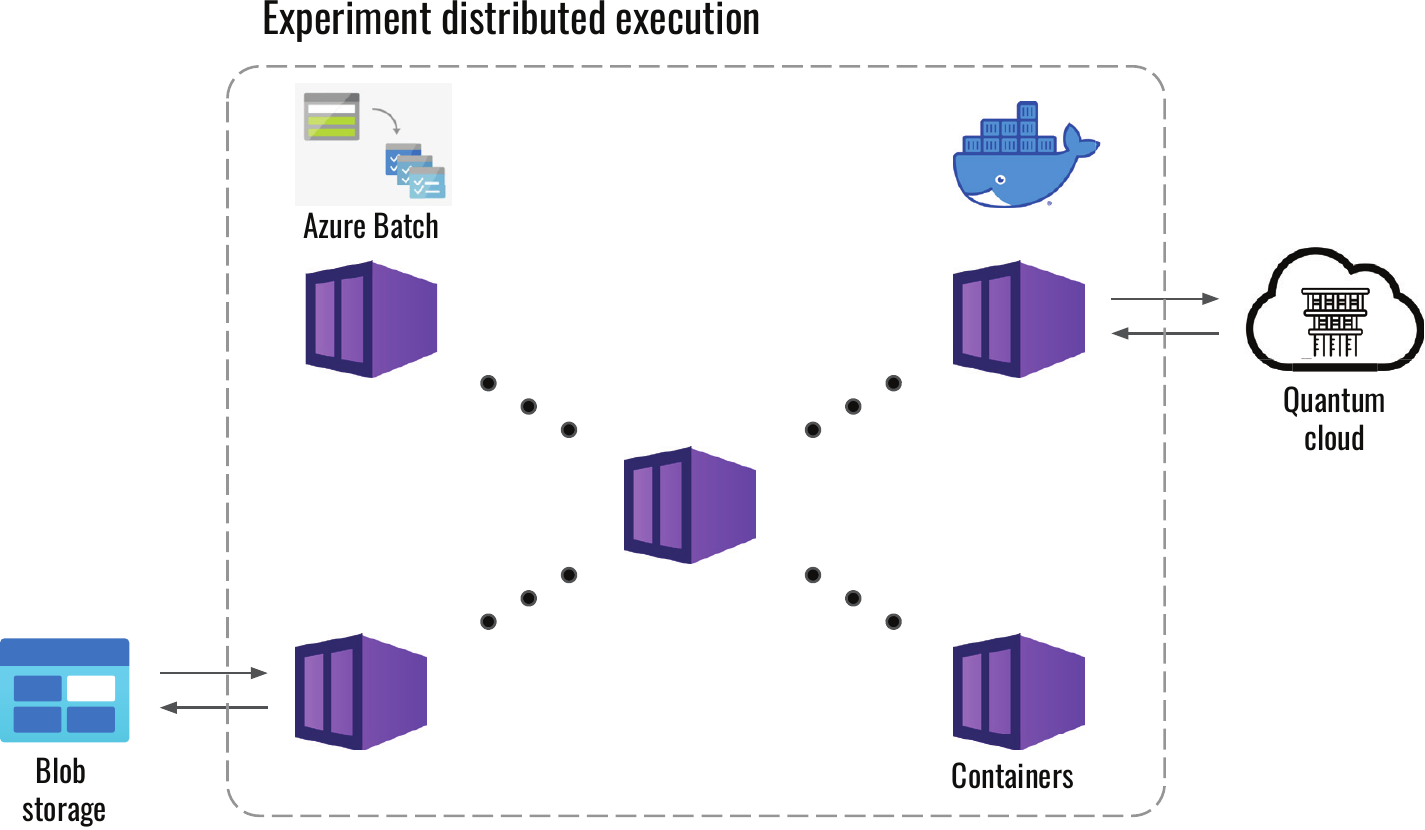}
    \caption{\textbf{Detailed view of the experiment execution pipeline} (Pipeline 1 in Fig. \ref{fig:overall_system_arch}). The process starts with Azure's Batch service spinning up the number of containers specified in the parameter search configuration file. The model combinations are distributed between the containers as evenly as possible. Each container runs its set of experiments independently, and on completion, persists the results to the central blob storage.}
    \label{fig:experiment_distributed_execution}
\end{figure}

The containers invoked by the three pipelines communicate through atomic reads and writes to a single object store, which in Azure is implemented via blob storage (see Fig. \ref{fig:experiment_distributed_execution}). File uploads to this storage can be used as a pipeline trigger in Data Factory. We rely on this feature to enforce the sequential execution of the pipelines, such that one stage only starts when the results of the previous one were persisted.

The initial trigger for the set of experiments is the upload of the parameter search configuration JSON file to blob storage. The second pipeline only begins upon successful completion of the first stage, and, when there are raw result files for as many nodes as defined in the experiment configuration.

Our current implementation distributes the experiments between the nodes following a static scheduling pattern: Before runtime, the total number of experiments is distributed between the available containers in sets of approximately equal size. Each container can communicate and submit jobs to quantum backends, such as the IBM quantum cloud.


\section{Example: Quantum Neural Networks}\label{sec:example_qnn}

In order to provide a concrete example of the developed framework described in Section \ref{sec:autoqml}, we focus on one particular type of quantum machine learning model -  the Quantum Neural Network (QNN). QNNs are a type of variational quantum circuit where the unitary gate operations are parametrized. Classical data is first encoded into a quantum state via pre-determined gate operations, and the resulting state is then fed into the parametrized circuit Ansatz \cite{zhao_review_2021}. The result is then measured via usual quantum measurement processes and transferred to a classical computer that computes a loss function. In this quantum formulation, the classical notion of backpropagation and standard weight/bias updates is implemented by updating the parameterized rotation gates relative to the result of the loss function calculation. By iteratively minimizing the loss function (for example using gradient-based optimizers such as ADAM or gradient-free methods as SPSA \cite{huggins_towards_2019}) and continuing to update the quantum circuit, the process is analogous to training a classical neural network. The attractiveness of QNNs lies in their potential to deliver training and inference speedups as well as their increased generalization capacity in specific scenarios \cite{abbas_power_2020}. However, how to accurately compare these benefits between classical and quantum neural networks is still an active area of research.

\subsection{Quantum GANs}
\label{section:quantum_gans}

One type of QNN that has been highly investigated both in theory and in terms of applied use-cases is the quantum Generative Adversarial Neural Network (qGAN) \cite{dallaire-demers_quantum_2018,lloyd_quantum_2018, romero_variational_2019}.

By leveraging quantum effects, these algorithms have the potential to outperform their classical counterparts by enabling quadratic speedups on model training, inference and expressive power. Additionally, their noise resilience makes them exceptionally well suited for running on NISQ devices \cite{huang_experimental_2021, zeng_learning_2019, gao_enhancing_2021}.

\begin{figure}[htb]
    \centering
    \includegraphics[width=0.5\textwidth]{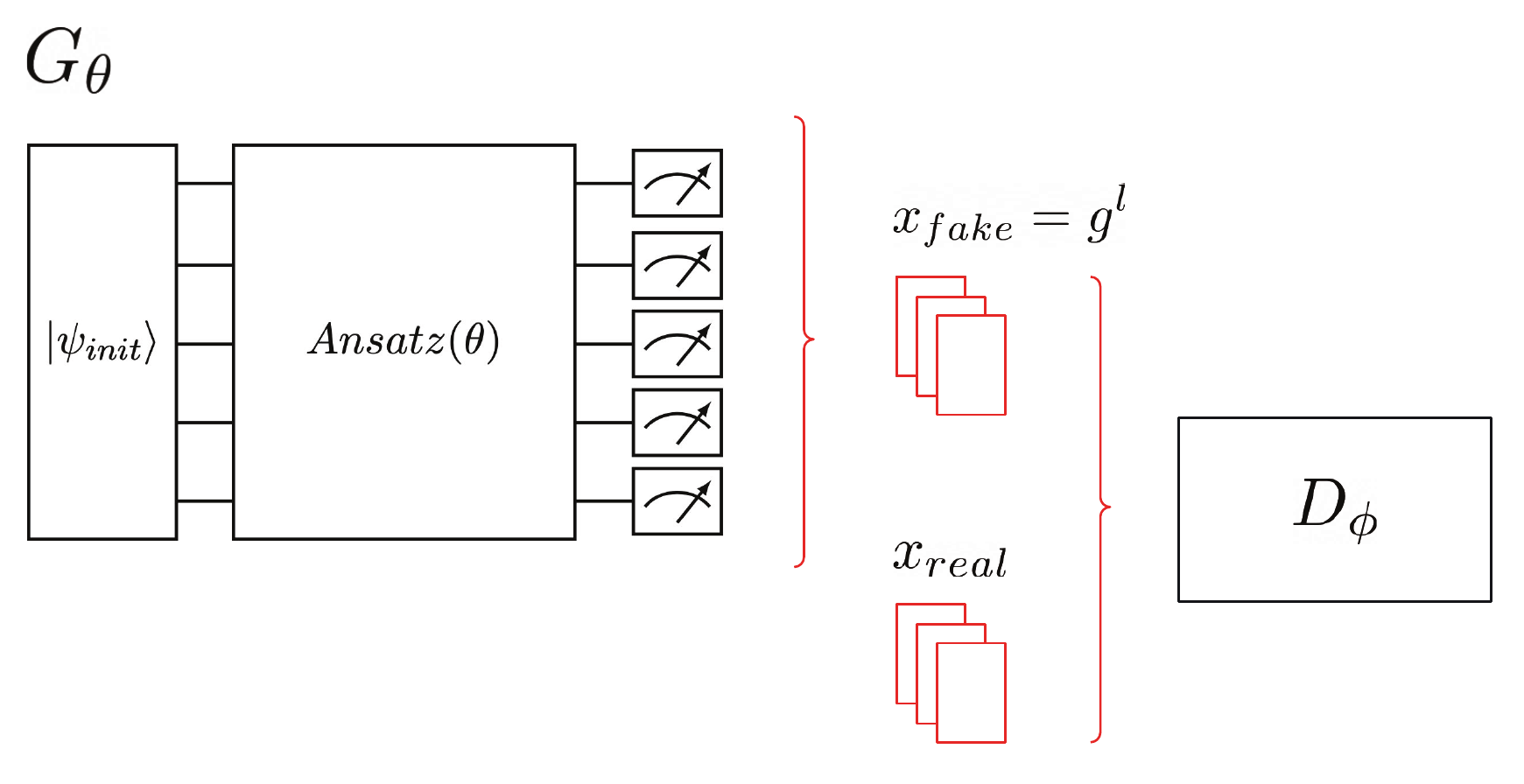}
    \caption{Generic architecture of a Born-sampled quantum GAN. The generator $G_{\theta}$ is a quantum circuit whose whole register is measured at once. The quantum state's bit strings are then mapped to the range of the data samples. The discriminator $D_{\phi}$ is a classical neural network.}
    \label{fig:qgan_architecture}
\end{figure}

We consider a generative model to be quantum whenever it uses quantum effects in its subroutines. In the case of GANs, the components that could be potentially quantum are the generator, the discriminator, and the data on which the model is trained. Here, quantum data refers to information whose natural representation is quantum. Fig. \ref{fig:qgan_architecture} describes the basic structure of the qGAN which has its generator implemented in a quantum circuit. The generator model is an arbitrary quantum Ansatz applied onto an initial state $|\psi_{\text{init}}\rangle$. There exist different methods of initial state choices and parameter initialization strategies. Our example QML training application will focus on Born-sampled qGANs with a classical discriminator \cite{zoufal_quantum_2019, situ_quantum_2020}. In this framework, we want to leverage quantum GANs to provide an efficient Qsample encoding of classical data in a quantum state. This qGAN is well suited to learn from floating-point, real-valued data, and it can be modified to learn complex distributions.

For the qGAN application, two important configuration choices need to be determined. First, the initial quantum state $|\psi_{\text{init}}\rangle$ which shall be used as input to the variational Ansatz needs to be determined. In \cite{zoufal_quantum_2019}, the authors provide evidence that an initial state closer to the target Qsample state will have better convergence properties and fit the target distribution better. Second, the actual choice of the Ansatz needs to be determined. 

The batches of samples are generated by measuring the qubits of the generator. Zoufal and Situ \cite{{zoufal_quantum_2019}} propose performing a global measurement of the qubit register, also known as Born sampling. In the Born-sampled case, the integer representation of the quantum states is mapped to the real range of the data following the proposed mapping:

\begin{equation}
    \phi(x) = a + \frac{b-a}{2^N-1}x \;,
\end{equation}
where $x \in \{0,...2^N-1\}$ are the integer representation of the basis states, $[a,b]$ is the value range of the distribution and $N$ is the number of generator qubits. The generated batch is combined with one of real samples and fed to the discriminator, which outputs the probability of samples being fake. These batch probabilities are used to compute the qGAN's loss function and its gradients.

\subsection{Model Result Metric}
\label{section:distribution-similarity}
Machine learning algorithms have a plethora of metrics used to measure the outcome of a model's result on a test set of unseen data. For our illustrative example of the framework which uses qGANs, measuring the degree of similarity between two distributions can be done in several standard ways. The primary metric we will use is the Kullback-Lieber divergence ($D_{KL}$) which measures the degree of similarity of two distributions. It is also known as relative entropy. Let $P$ and $Q$ be discrete probability distributions defined in the probability space $\mathcal{X}$. Then the KL-divergence (relative entropy) is given by

\begin{equation}
    D_{\text{KL}}(P||Q) = \sum_{x\in\mathcal{X}}P(x)\log\left(\frac{P(x)}{Q(x)}\right)\;,
\end{equation}
where $P$ is the target distribution and $Q$ is the model. 

\section{Application: Auto Training qGAN for Energy Price Distribution}\label{sec:application}

This section applies the automation framework for a particular machine learning task that has applications to energy economics option pricing. In \cite{zoufal_quantum_2019}, the authors demonstrated that one can use $\mathcal{O}(poly(n))$ number of gates as opposed to $\mathcal{O}(n^2)$ gates for efficiently loading a generic random distribution into a quantum circuit using a qGAN. Using this advantage, they go on to implement an algorithm that uses quantum amplitude estimation to provide a quadratic runtime improvement for 
European call option pricing over stand Monte Carlo techniques \cite{stamatopoulos_option_2020}.

In the energy trading sector, one typical data source which is used is the open-market energy prices. Depending on the country and market, energy prices can be traded and fluctuate in 15-minute intervals to guarantee that supply, demand, and network congestion match. Human consumption patterns lead to seasonal, daily, and hourly trends. The upper histogram in Fig. \ref{fig:best_qgan_result} represents the normalized distribution of energy prices in the year 2015 obtained from the EPEX SPOT market between the hours of 00:00-01:00. The distribution is non-trivial and will serve as an illustrative example of our framework. The goal is to train a qGAN to efficiently load the distribution so that we may use it in future energy economics quantum algorithms.

\subsection{Deployment Configuration}

In order to run the automated training framework, one must create a suitable configuration file in JSON format. This central configuration file defines machine learning hyperparameters and quantum-specific parameters such as Ansatz type, number of Ansatz repetitions, and quantum hardware backend. An example configuration file is given below:
\medskip
\begin{lstlisting}[language=json,firstnumber=1,basicstyle=\small,xleftmargin=5.0ex]
{
  "name": "qGAN fitting the E.ON pricing data using different Ansaetze",
  "goal": "Test the capacity of a each circuit",
  "metrics": "relative_entropy",
  "n_containers": 10,
  "visualizations": ["entanglement_histogram", "entropy_curve"],
  "distributions": [
    {
      "data_path": "\qGAN\data\eon_midnight.csv",
      "samples": 20000,
      "discretization": "optimal"
    }],
  "ansaetze": [
    {
      "type": "zoufal",
      "repetitions": [1,2,3]
    },
    {
      "type": "vallecorsa",
      "repetitions": [1,2,3]
    },
    {
      "type": "herr_1",
      "repetitions": [1,2,3]
    }],
  "initializations": [{ "type": "uniform" }, { "type": "normal" }, { "type": "Random" }],
  "num_qubits": [2, 3, 4, 5, 6],
  "batch_size": 512,
  "num_epochs": 2000,
  "num_training_runs": 10,
  "discriminator": {
    "type": "custom_classical_1",
    "hparams": {
      "lr": [1e-4],
      "n_hidden": [20],
      "n_input": 50,
      "betas": [0.9, 0.999]
    },
    "type": "custom_classical_2",
    "hparams": {
      "lr": [1e-4],
      "n_hidden": [40,10],
      "n_input": 50,
      "betas": [0.9, 0.999]
    },
  },
  "optimizer": {
    "lr": [1e-3,1e-4],
    "betas": [0.7, 0.99]
  }}
\end{lstlisting}

In this sample configuration file, we capture hundreds of possible qGAN models. Notably, we use existing qGAN architectures from \cite{zoufal_quantum_2019,herr_anomaly_2020} and \cite{chang_dual-parameterized_2021}, and, for each circuit Ansatz we train models with multiple layers (1, 2 or 3 applications) along with two different classical discriminator architectures, $custom\_classical\_1$ and $custom\_classical\_2$. We also implement three different quantum state initialization strategies (uniform, normal or random) and use the relative entropy as the main performance metric. Due to the large runtime overhead, we experimentally found that for our use case, 10 runs per configuration was an appropriate amount to obtain results based on the mean and variance of the outcomes. The next section discusses some sample results where Table \ref{tab:5-opt-ibm} 
presents the statistical outcomes of $n=10$ training runs from the 5-qubit fitted distribution.
\subsection{Results}

Our framework is set up to detect new training configuration files uploaded to the cloud-based storage account. This event triggers a pipeline that initiates the classical-quantum parallel training as described in Fig. \ref{fig:experiment_distributed_execution}. All numeric and plotting results are stored in blob storage, while the models are stored in Python's pickle format to enable more effortless model loading.

\begin{figure}[!hb]
    \centering
    \begin{minipage}[b]{0.46\textwidth}
    \includegraphics[width=\textwidth]{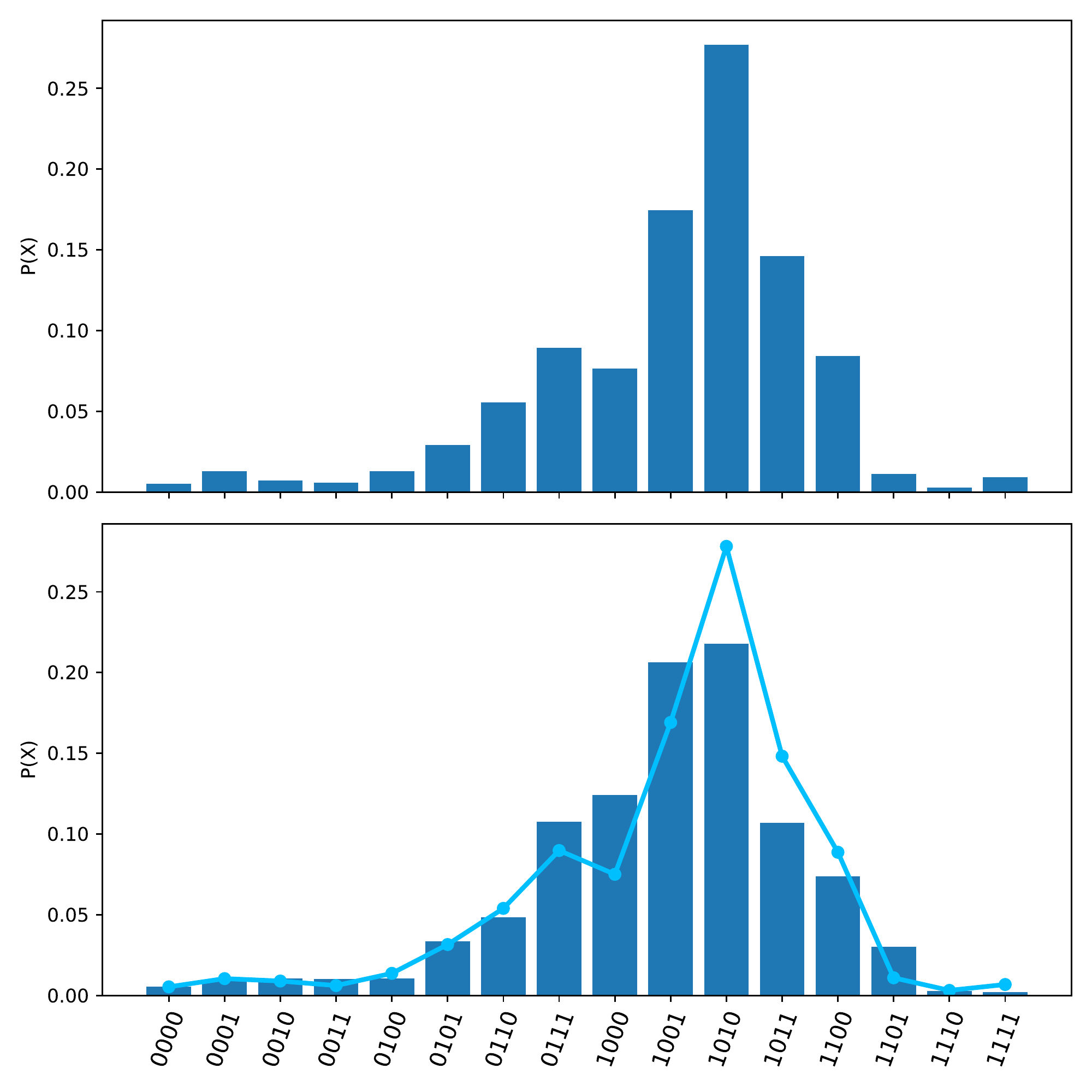}
    \end{minipage}
    \caption{Top: Target distribution of energy price data. Historic hourly SPOT prices from EPEX market for 2015 were collected and averaged per hour. Using our automated training framework, we find the best qGAN network which reproduces this target distribution. Bottom: The output from the best qGAN selected by the automated training framework factoring in the Relative Entropy, Kolmogorov-Smirnov test, and training loss functions. Target distribution from the top figure is displayed as light blue and resulting quantum generated histogram in dark blue.}
    \label{fig:best_qgan_result}
\end{figure}

Table \ref{tab:5-opt-ibm} collects the statistic results of an experiment with a 5-qubit Zoufal Ansatz for different combinations of initialization strategies and an increasing number of layers. In the table; $k$ is the number of Ansatz application layers, $\mu_{\text{KS}}$ and $\sigma_{\text{KS}}$ are Kolmogorov-Smirnov measures of output distribution compared to the target distribution for 10 training runs, $\mu_{\text{RE}}$ and $\sigma_{\text{RE}}$ are the same metrics but using the relative entropy (Kullback-Lieber Divergence) similarity metric, and $\mu_{\text{Depth}}$ is the mean and standard deviation of gate depth after transpilation to IBM Quantum hardware. The best 5 qubit model selected is the uniform 2-layer with gate depth 77 (bolded).

\begin{table}[!h]
    \renewcommand{\arraystretch}{2.5}
            \caption{Fitting energy price distribution with the 5-Qubit Zoufal Ansatz}

    \begin{tabular}{cccccccc} 
              \textbf{Initialization} & \textbf{k}& \textbf{$\mu_{\text{KS}}$} & \textbf{$\sigma_{\text{KS}}$} & \textbf{$\mu_{\text{RE}}$} & \textbf{$\sigma_{\text{RE}}$} & \textbf{$\mu_{Depth}$}  \\ 
             \hline
                 \textbf{Uniform} &1& 0.1780  & 0.0519 & 0.5692 &  0.1325 & 41.18 \\[-4pt] 
                ~&\textbf{2}& \textbf{0.1104}  & \textbf{0.0470} & \textbf{0.3562} &  \textbf{0.0947} & \textbf{77.09} \\
                ~&3& 0.1540  & 0.0807 & 0.4329 & 0.2479 & 104.52 \\
                 Normal & 1& 0.1570  & 0.0389 & 0.2793 &  	0.0269  & 203.42 \\[-4pt] 
                ~ & 2& 0.1446  & 0.0531 & 0.2434 &  0.0383  & 238.38 \\
                ~&3& 0.1516  & 0.0305 & 0.2510 & 0.0343 & 271.2 \\
                Random &1&  0.3420  & 0.1676 &1.1412 &  0.6072 & 33.75 \\
               ~ &2&  0.1992  & 0.0970 &0.7595 &  0.3290	 & 74.55 \\
                ~&3& 0.1536  & 	0.1034 & 0.5494 & 0.4724 & 101.89\\
        \end{tabular} 
        \label{tab:5-opt-ibm}
\end{table}

\begin{figure*}[htb]
    \centering
    \includegraphics[width=1.0\textwidth]{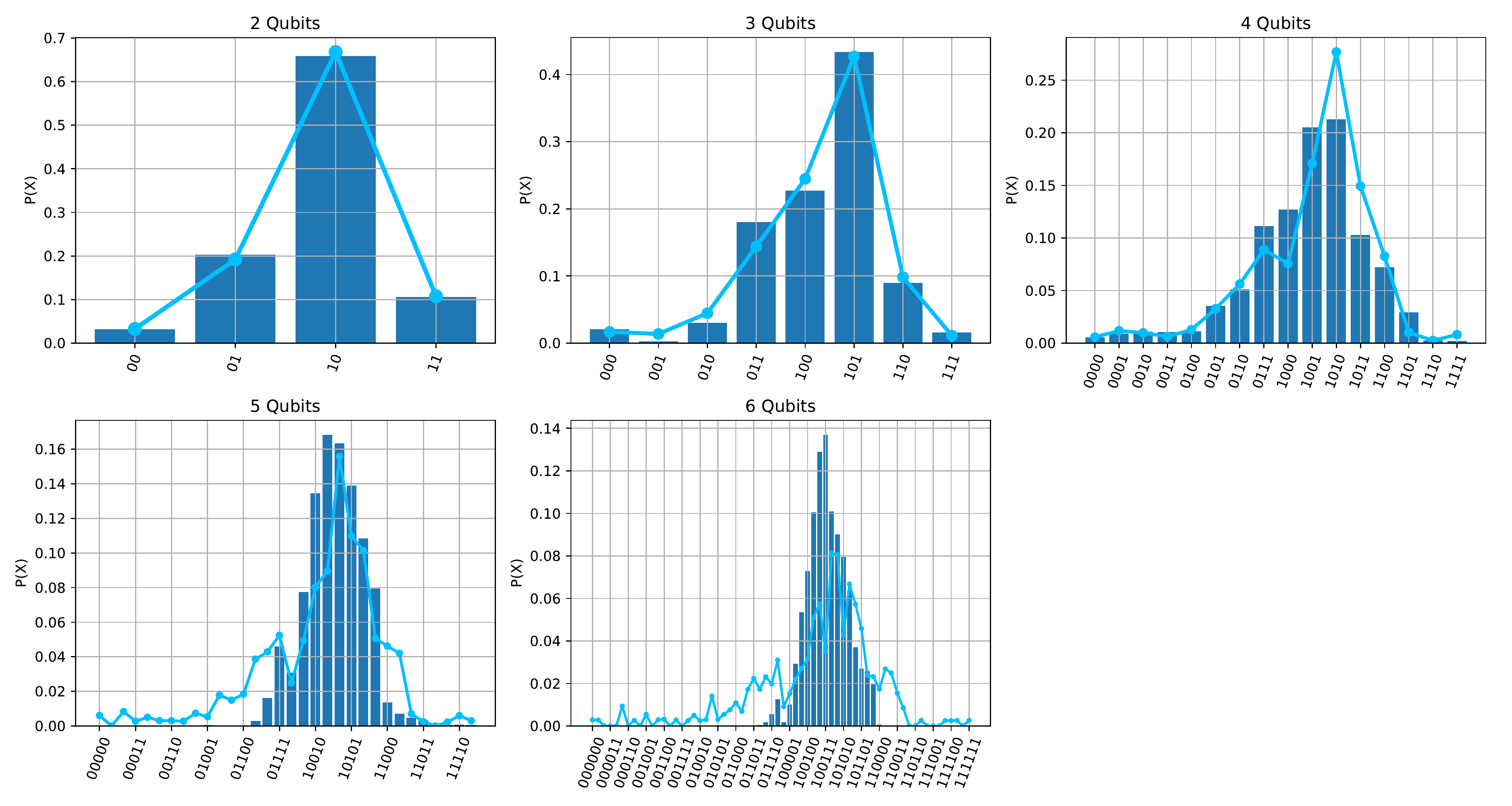}
    \caption{Sample output of the tool demonstrating the qubit scalability of certain Ansatz for qGAN learning of the target distribution. Original data is displayed as light blue, and resulting quantum generated histogram is dark blue.}
    \label{fig:qgan_scalability_output}
\end{figure*}

\begin{figure}[!hb]
  \centering
  \begin{minipage}[b]{0.40\textwidth}
    \includegraphics[width=\textwidth]{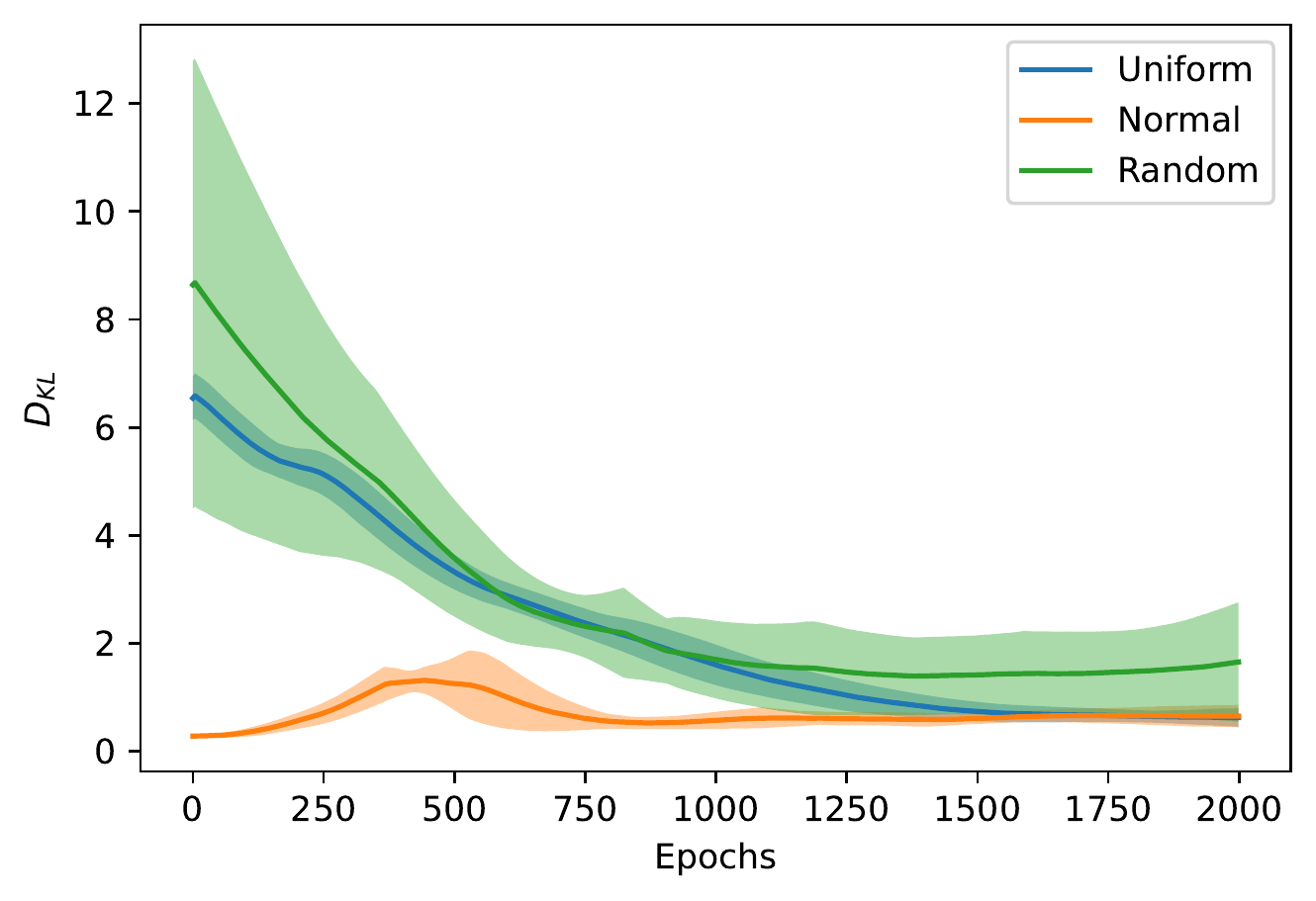}
  \end{minipage}
  \hfill
  \begin{minipage}[b]{0.41\textwidth}
    \includegraphics[width=\textwidth]{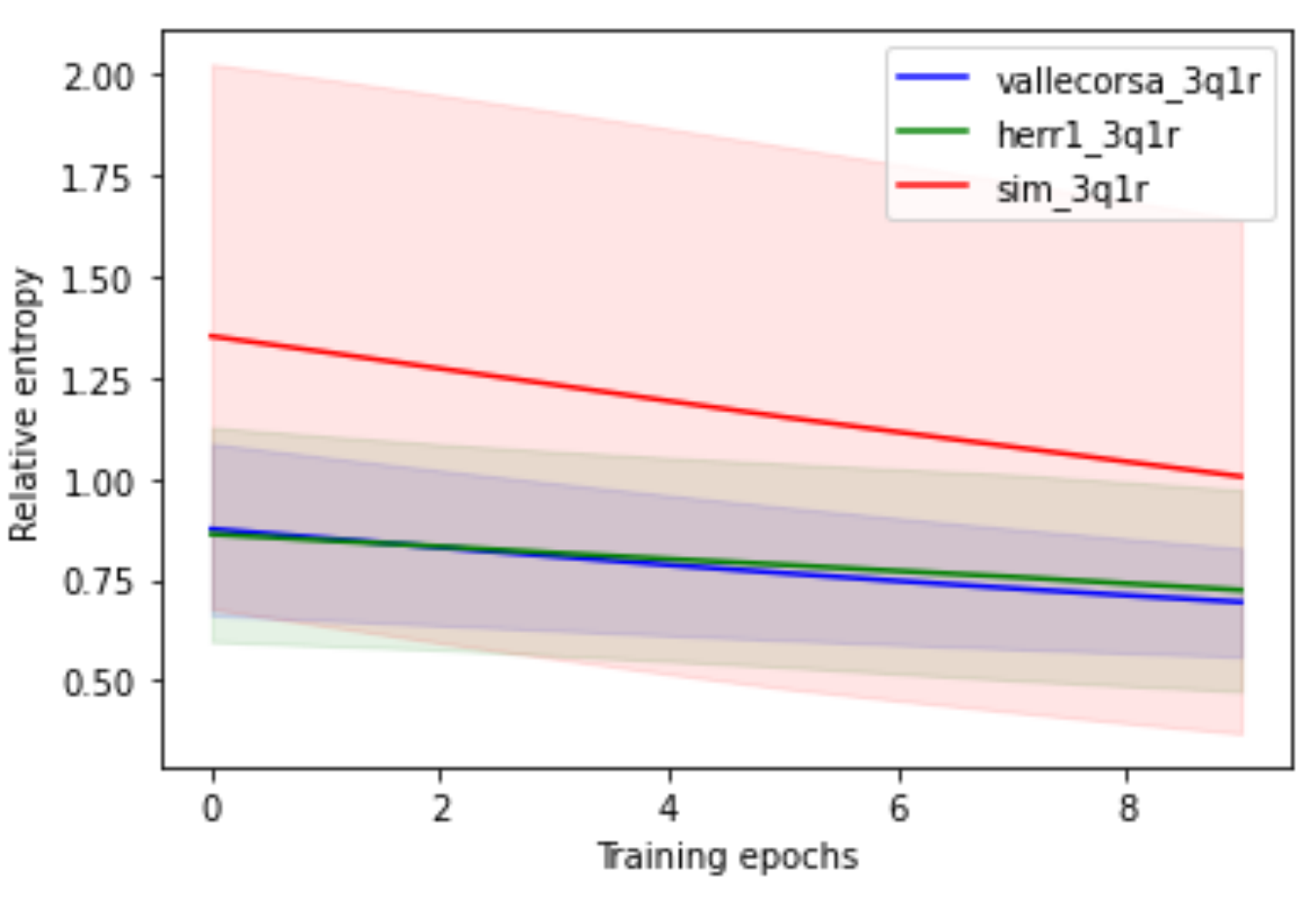}
  \end{minipage}
  \hfill
    \begin{minipage}[b]{0.41\textwidth}
    \includegraphics[width=\textwidth]{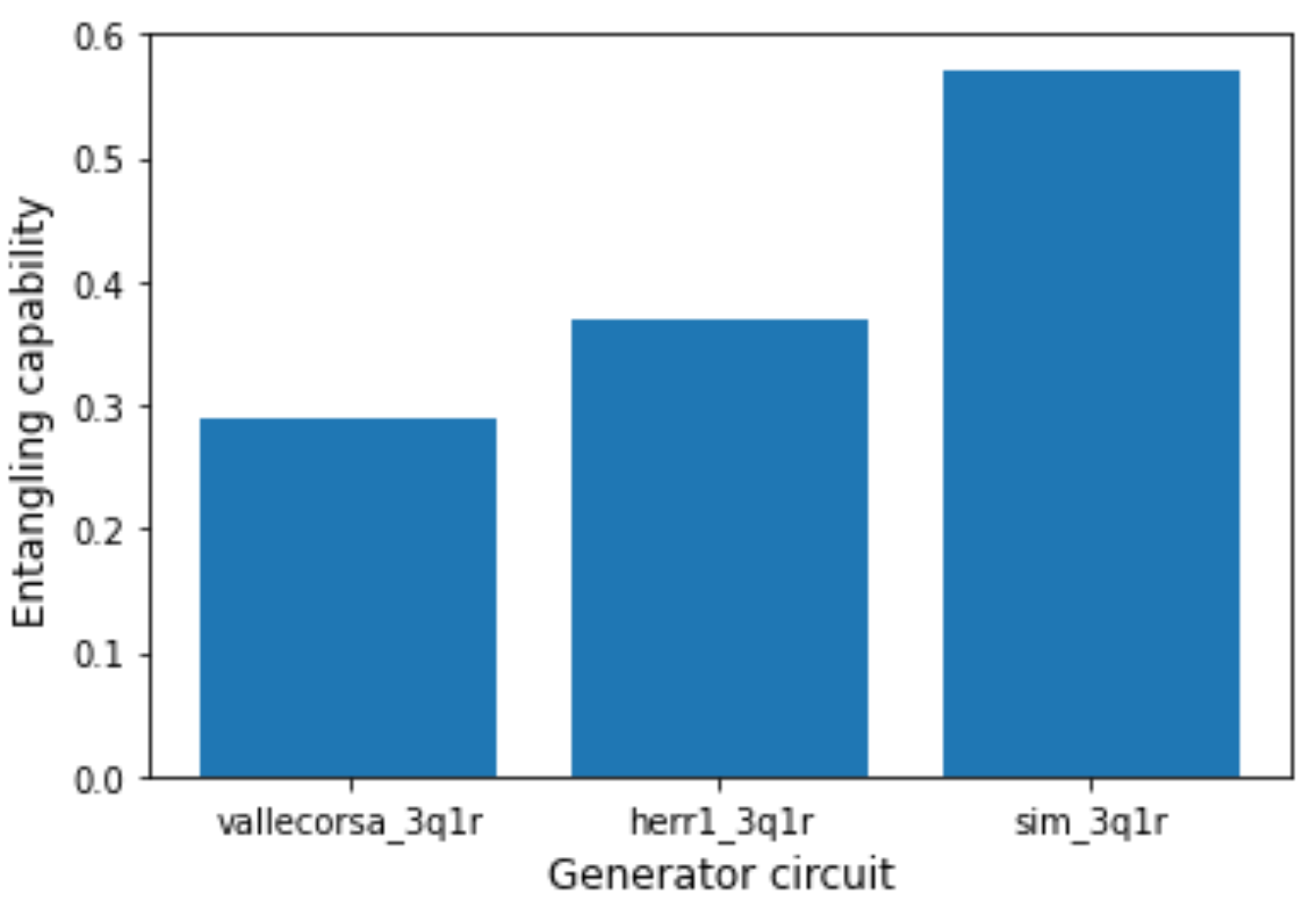}
    \end{minipage}
      \caption{Sample results output of the automated parameter tuning tool for training qGANs to fit the target distribution. 
      Top: Entropy curves of the IBM Zoufal qGAN Ansatz for different initialization strategies (Uniform, normal and random). Middle: Entropy curves for different qGAN Ansatz. 
      Bottom: Entangling capacity of different Ansatz which are being optimized over.}
      \label{fig:sample_framework_output_plots}
\end{figure}

These qGAN outputs are given in Fig. \ref{fig:qgan_scalability_output}, with the final best model as determined by the framework demonstrated in Fig. \ref{fig:best_qgan_result}. Fig. \ref{fig:sample_framework_output_plots} shows a selection of other useful output data which were generated for model comparison. Details such as training loss curves as a function of initialization strategy, circuit Ansatz type, and the entangling capacity for a given circuit are useful for the quantum data engineer to verify that the automation task was a success.

\section{Conclusion}\label{sec:conclusion}
In this paper, we introduce the notion of Automated Quantum Machine Learning (AutoQML) and emphasize its importance in the future of quantum machine learning.
As the first steps towards the solution of the proposed problem, we developed an overarching classical-quantum distributed model training execution framework. From the quantum data engineering perspective, it works by defining an efficient configuration object structure which can be remote deployed to a cloud platform where QML jobs can then run in parallel. The best model is then returned as the final solution. The framework described can be applied to any QML algorithm such as quantum neural networks, kernel methods, clustering, and classification algorithms. 

As a first use case of the framework, we apply it for quantum Generative Adversarial Network (qGANs) training for solving a data-loading problem relevant in the energy trading sector. 

In implementing the first version of such a solution, we discovered several exciting avenues to explore. The most obvious is the generalization of our model selection algorithm. The brute-force combinatorial approach to model parallelization and training should be simplified. See, for example, the Bayesian optimization techniques which were recently introduced into the standard Python data science library scikit-learn to create Auto-sklearn \cite{feurer2019auto,feurer2021auto}. We envision that similar add-ons and optimization techniques can be included in our solution architecture in future versions.

Another model-specific improvement that could be made is the ability to optimize the form of circuit Ansatz (the choice of unitary gates) chosen for a particular machine learning task. Initial steps have been made in this direction for the qGAN data loading problem and will be discussed in future work. In our example use-case, we tested several existing circuit architectures, however, as in classical automated machine learning for neural networks, the ability to automatically determine the specific layer \emph{and} node (gate) types or counts from scratch will be the future of this application.

In terms of quantum hardware architecture, one can envision the system running QPU-specific analysis, taking into account different qubit connectivities/topologies, chip types/generations, and circuit transpilations that reduce to the specific operational basis gates depending on the exact chip architecture. Certainly, different machine learning models will perform better or worse on newer, more state-of-the-art devices with better hardware-specific properties such as general coherence times, quantum volume, qubit count, qubit connectivity and Circuit Layer Operations Per Second (CLOPS).

From the software architecture and classical cloud perspective, many open topics can also be further investigated. For one, the proposed architecture may generalize to allow for single problems to be run in parallel on multiple QPU devices. In our initial implementation, each classical container can remote-submit jobs individually to quantum hardware. However, one can imagine performing batched process deployment across different quantum hardware as NISQ devices increase in their capacities and availabilities. This could allow for even more parallelization to explore potential model space. 

Another cloud-based solution one might consider is the creation of a Quantum Model Store (QMS). In such a version-controlled environment, the resulting optimized models are stored and used for runtime predictions. Standardized solutions will be necessary when hybrid classical-quantum software architectures become a reality in industry. These solutions will be either developed in-house or offered by cloud providers akin to Microsofts' AzureML or Amazon Web Services' Sagemaker. 

As quantum machine learning algorithms mature and quantum hardware continues to scale, we expect these types of software architectural patterns to continue developing. In classical machine learning, Automated ML solutions and cloud based machine learning workflows are now mainstream in industry. Cloud providers such as Google, Microsoft and AWS already have platforms for offering access to quantum hardware, so it is a natural extension that AutoQML and model management be adopted in the future.
\vspace{200px}
\nocite{qc_icons}

\end{document}